\documentclass[12pt]{article}

\newcommand{\etal}{{\em et al.}}
\newcommand{\ie}{{\it i.e.}}
\newcommand{\ket}[1]{\vert\,{#1}\rangle}
\newcommand{\VEV}[1]{\left\langle{#1}\right\rangle}
\renewcommand{\bar}[1]{\overline{#1}}
\newcommand{\M}{{\cal M}}
\newcommand{\eg}{{\it e.g.}}

\begin {document}
\begin{flushright}
{\small
SLAC--PUB--8361\\
February 2000\\}
\end{flushright}

\begin{center}
{{\bf\LARGE   
Hadronic Light-Cone Wavefunctions and the Unification of
QCD Bound-State Phenomena}\footnote{Work supported by
Department of Energy contract  DE--AC03--76SF00515.}}

\bigskip
Stanley J. Brodsky\\
{\sl Stanford Linear Accelerator Center \\
Stanford University, Stanford, California 94309}
\end{center}

\vfill

\begin{center}
{\bf\large   
Abstract }
\end{center}

The light-cone Fock representation encodes
the bound-state quark and gluon properties of hadrons, including their
helicity and flavor correlations, in terms of universal
process-independent and frame-independent wavefunctions.  It also provides
a physical factorization scheme for separating hard and soft contributions
in both exclusive and inclusive hard processes.  A new type of jet
production reaction, ``self-resolving diffractive interactions" can
provide direct information on the light-cone wavefunctions of hadrons in
terms of their QCD degrees of freedom as well as the composition of
nuclei in terms of their nucleon and mesonic degrees of freedom.

\vfill

\begin{center} 
{\it Invited talk presented at} \\
{\it Circum-Pan-Pacific RIKEN Symposium on} \\
{\it ``High Energy Spin Physics ''} \\
{\it RIKEN, Wako, Japan }\\
{\it November 3--6, 1999}\\
\end{center}

\vfill

\section{Introduction}

Ever since the discovery of the Bjorken scaling of deep inelastic
lepton-proton scattering in 1967, \cite{Coward:1968au} high energy
lepton scattering experiments have provided increasingly detailed
information on the flavor, momentum, and helicity distributions of the
quarks and gluons in hadrons.  The results are represented
in the form of leading-twist light-cone momentum and helicity
distributions
$q(x,\lambda, Q)$, $\bar q(x,\lambda, Q)$ and $g(x, \lambda, Q)$ at the
resolution of  $Q$.  However, such distributions
represent single-particle probabilities and thus do not contain
information on the transverse momentum, spin, and flavor correlations of
the bound quarks and gluons.  In particular, structure functions cannot
specify the phases needed to understand QCD processes at the amplitude
level, the physics which underlies form factors, exclusive and
diffractive scattering processes, and the hadronic decay amplitudes of
heavy hadrons.  The polarized beam and polarized target experiments now in
progress and planned at Jefferson Laboratory, HERMES at DESY, BNL, CERN,
and SLAC, and measurements of rare exclusive channels
and their polarization correlations in
$e^+ e^-$ and $\gamma \gamma$ annihilation at the high luminosity $B$
factories promise a new level of precision in testing QCD and determining
fundamental properties of hadrons.  A global unified interpretation of
such inclusive and exclusive experiments is a challenging theoretical
problem, mixing issues involving non-perturbative and perturbative
dynamics.

Ideally, one wants to have a frame-independent, quantum-mechanical
description of had\-rons at the amplitude level capable of
encoding all possible quark and gluon momentum, helicity, and flavor
correlations in the form of universal process-independent hadron
wavefunctions for each particle number configuration.
Remarkably, the light-cone Fock expansion allows just such a unifying
representation.  Moreover, the light-cone
formalism provides a physical factorization scheme which conveniently
separates and factorizes soft non-perturbative physics from hard
perturbative dynamics in both exclusive and inclusive reactions.

Formally, the light-cone expansion is constructed by quantizing QCD at
fixed light-cone time\cite{Dirac:1949cp} $\tau = t + z/c$ and forming the
invariant light-cone Hamiltonian: $ H^{QCD}_{LC} = P^+ P^- - P^2_\perp$
where
$P^\pm = P^0 \pm P^z.$\cite{PinskyPauli} The momentum
generators
$P^+$ and
$P_\perp$ are kinematic; \ie, they are independent of the interactions.
The generator
$P^- = i {d\over d\tau}$ generates light-cone time translations, and
the eigen-spectrum of the Lorentz scalar $ H^{QCD}_{LC}$ gives the
mass spectrum of the color-singlet hadron states in QCD together with
their respective light-cone wavefunctions.  For example, the
proton state satisfies:
$ H^{QCD}_{LC} \ket{\psi_p} = M^2_p \ket{\psi_p}$.  The expansion of
the proton eigensolution $\ket{\psi_p}$ on the color-singlet
$B = 1, Q = 1$ eigen states $\{\ket{n} \}$
of the free Hamiltonian $ H^{QCD}_{LC}(g = 0)$ gives the
light-cone Fock expansion:
$ \vert \psi_p(P^+, P_\perp )> = \sum_n \psi_n(x_i, k_{\perp i},
\lambda_i) \vert n; x_i P^+, x_i P_\perp + k_{\perp i}, \lambda_i>$.

The light-cone momentum fractions
$x_i = k^+_i/P^+$ with $\sum^n_{i=1} x_i = 1$ and $k_{\perp i}$ with
$\sum^n_{i=1} k_{\perp i} = 0_\perp$ represent the relative momentum
coordinates of the QCD constituents.  The physical transverse momenta are
$p_{\perp i} = x_i P_\perp + k_{\perp i}.$ The
$\lambda_i$ label the light-cone spin $S_z$ projections of the quarks and
gluons along the quantization $z$ direction.  The physical gluon
polarization vectors
$\epsilon^\mu(k, \lambda = \pm 1)$ are specified in light cone
gauge $k \cdot \epsilon = 0, \eta \cdot \epsilon = \epsilon^+ = 0.$
Light-cone quantization is most conveniently carried out in the physical
ghost-free light-cone gauge $A^+ = 0;$ however,
light-cone quantization in Feynman gauge also has a number of attractive
features, including manifest covariance and a straightforward passage to
the Coulomb limit in the case of heavy static quarks.\cite{Srivastava:1999gi}

The solutions of $ H^{QCD}_{LC} \ket{\psi_p} = M^2_p \ket{\psi_p}$
are independent of $P^+$ and $P_\perp$;  thus given the
eigensolution Fock projections $ \VEV{n; x_i, k_{\perp i},
\lambda_i \vert p} = \psi_n(x_i, k_{\perp i},
\lambda_i) ,$ the wavefunction of the proton is determined in any
frame.\cite{LB}
In contrast, in equal-time quantization,  a Lorentz boost
always mixes dynamically with the interactions, so that computing a
wavefunction in a new frame requires solving a nonperturbative problem
as complicated as the Hamiltonian eigenvalue problem itself.

The LC wavefunctions $\psi_{n/H}(x_i,\vec
k_{\perp i},\lambda_i)$ are universal, process independent, and thus
control all hadronic reactions.
Given the light-cone wavefunctions, one can compute the
moments of the helicity and
transversity distributions measurable in polarized deep inelastic
experiments.
Similarly, the matrix elements of the currents as
integrated squares of the LC wavefunctions. \cite{LB} For example, the
polarized quark distributions at resolution $\Lambda$ correspond to
\begin{eqnarray}
q_{\lambda_q/\Lambda_p}(x, \Lambda)
&=& \sum_{n,q_a}
\int\prod^n_{j=1} dx_j d^2 k_{\perp j}\sum_{\lambda_i}
\vert \psi^{(\Lambda)}_{n/H}(x_i,\vec k_{\perp i},\lambda_i)\vert^2
\\
&& \times \delta\left(1- \sum^n_i x_i\right) \delta^{(2)}
\left(\sum^n_i \vec k_{\perp i}\right)
\delta(x - x_q) \delta_{\lambda_a, \lambda_q}
\Theta(\Lambda^2 - {\cal M}^2_n) \nonumber
\end{eqnarray}
where the sum is over all quarks $q_a$ which match the quantum
numbers, light-cone momentum fraction $x,$ and helicity of the struck
quark.  Similarly, moments of transversity distributions and other
off-diagonal helicity convolutions are defined as a density matrix of the
light-cone wavefunctions.  The light-cone wavefunctions
also specify the multi-quark and gluon correlations of the hadron.  For
example,  the distribution of spectator particles in the final state
which could be measured in the proton fragmentation region in deep
inelastic scattering at an electron-proton collider are in principle
encoded in the light-cone wavefunctions.

The effective
lifetime of each configuration in the laboratory frame is
$2 P_{\rm lab}/$ $({\M}_n^2- M_p^2) $
where
$ \M^2_n = \sum^n_{i=1}(k^2_{\perp i} + m^2_i)/x_i < \Lambda^2 $
is the off-shell invariant mass and $\Lambda$ is a global
ultraviolet regulator.
The light-cone momentum integrals are thus be limited by requiring that
the invariant mass squared of the constituents of each Fock state is less
than the resolution scale $\Lambda$.  As I discuss below, this cutoff
serves to define a factorization scheme for separating hard and soft
regimes in both exclusive and inclusive hard scattering
reactions.\cite{LB}

The ensemble {$\psi_{n/H}$} of light-cone Fock
wavefunctions is a key concept for hadronic physics, providing the
interpolation between physical hadrons (and also nuclei)
and their fundamental quark and gluon degrees of freedom.  Each
Fock state interacts distinctly; \eg\ Fock states with small particle
number and small impact separation have small color dipole moments and can
traverse a nucleus with minimal interactions.  This is the basis for the
predictions for ``color transparency".  \cite{BM}

Given the
$\psi^{(\Lambda)}_{n/H},$ one can construct any spacelike electromagnetic
or electroweak form factor or local operator product matrix element from
the diagonal overlap of the LC wavefunctions.\cite{BD} Similar results
hold for the matrix elements which occur in deeply virtual Compton
scattering.  Exclusive semi-leptonic
$B$-decay amplitudes such as $B\rightarrow A \ell \bar{\nu}$ can also be
evaluated exactly.\cite{Brodsky:1998hn}
In this case, the timelike decay matrix elements require the
computation of both the diagonal matrix element $n \rightarrow n$ where
parton number is conserved and the off-diagonal $n+1\rightarrow n-1$
convolution such that the current operator annihilates a $q{\bar{q'}}$
pair in the initial $B$ wavefunction.  This term is a consequence of the
fact that the time-like decay $q^2 = (p_\ell + p_{\bar{\nu}} )^2 > 0$
requires a positive light-cone momentum fraction $q^+ > 0$.  Conversely
for space-like currents, one can choose $q^+=0$, as in the
Drell-Yan-West representation of the space-like electromagnetic form
factors.\cite{BD} However, as can be seen from the explicit analysis of
timelike form factors in a perturbative model, the off-diagonal
convolution can yield a nonzero $q^+/q^+$ limiting form as $q^+
\rightarrow 0$.  This extra term appears specifically in the case of
``bad" currents such as
$J^-$ in which the coupling to $q\bar q$ fluctuations in the light-cone
wavefunctions are favored.  In effect, the $q^+ \rightarrow 0$ limit
generates $\delta(x)$ contributions as residues of the $n+1\rightarrow
n-1$ contributions.  The necessity for such ``zero mode" $\delta(x)$ terms
has been noted by Chang, Root and Yan \cite{CRY}, Burkardt \cite{BUR},
and Ji and Choi.\cite{Choi:1998nf}

The off-diagonal $n+1 \rightarrow n-1$ contributions give a new
perspective for the physics of $B$-decays.  A semi-leptonic decay
involves not only matrix elements where a quark changes flavor, but also
a contribution where the leptonic pair is created from the annihilation
of a $q {\bar{q'}}$ pair within the Fock states of the initial $B$
wavefunction.  The semi-leptonic decay thus can occur from the
annihilation of a nonvalence quark-antiquark pair in the initial hadron.
This feature carries over to exclusive hadronic $B$-decays, such as
$B^0 \rightarrow \pi^-D^+$.  In this case the pion can be produced from
the coalescence of a $d\bar u$ pair emerging from the initial higher
particle number Fock wavefunction of the $B$.  The $D$ meson is then
formed from the remaining quarks after the internal exchange of a $W$
boson.

Light-cone Fock state wavefunctions thus encode all of the bound state
quark and gluon properties of hadrons such as spin and flavor
correlations in the form of universal process- and frame- independent
amplitudes.  Is there any hope of computing these wavefunctions from
first principles?  In the discretized light-cone quantization
method (DLCQ), \cite{Pauli:1985ps}
periodic boundary conditions are introduced in
$b_\perp$ and $x^-$ so that the momenta
$k_{\perp i} = n_\perp \pi/ L_\perp$ and $x^+_i = n_i/K$ are
discrete.  A global cutoff in invariant mass of the partons in the Fock
expansion is also introduced.
Solving the quantum field theory then reduces to
the problem of diagonalizing the finite-dimensional hermitian matrix
$H_{LC}$ on a finite discrete Fock basis.  The DLCQ method has now become
a standard tool for solving both the spectrum and light-cone wavefunctions
of one-space one-time theories.  Virtually any
$1+1$ quantum field theory, including ``reduced QCD" (which has both quark and
gluonic degrees of freedom) can be completely solved using
DLCQ.\cite{Kleb,AD} Hiller, McCartor, and I
\cite{Brodsky:1998hs,Brodsky:1999xj} have recently shown that the use of
covariant Pauli-Villars regularization with discrete light-cone
quantization allows one to obtain the spectrum and light-cone
wavefunctions of simplified theories in physical space-time dimensions,
such as (3+1) Yukawa theory.  Dalley \etal\ have also showed how one can
use DLCQ with a transverse lattice to solve
gluonic QCD.\cite{Dalley:1999ii}
Remarkably, the spectrum obtained for gluonium states is in remarkable
agreement with lattice gauge theory results, but with a huge reduction of
numerical effort.  One can also formulate DLCQ so that supersymmetry is
exactly preserved in the discrete approximation, thus combining the power
of DLCQ with the beauty of
supersymmetry.\cite{Lunin:1999ib,Haney:1999tk} The ``SDLCQ" method has
been applied to several interesting supersymmetric theories, to the
analysis of zero modes, vacuum degeneracy, massless states, mass gaps, and
theories in higher dimensions, and even tests of the
Maldacena conjecture.\cite{Antonuccio:1999ia}
Broken supersymmetry is interesting in DLCQ, since it may serve as a
method for regulating non-Abelian theories. \cite {Brodsky:1999xj}

Another
remarkable advantage of light-cone quantization is that the vacuum state
$\ket{0}$ of the full QCD Hamiltonian coincides with the free vacuum.
For example, as discussed by Bassetto,\cite{Bassetto:1999tm}
the computation of the spectrum
of $QCD(1+1)$ in equal time quantization requires constructing the full
spectrum of non perturbative contributions (instantons).  However,
light-cone methods such as DLCQ, give the correct result immediately,
without any need for vacuum related contributions.

It is also possible to model the light-cone wavefunctions.  For example
one can find simple forms for the three valence quark wavefunctions
$\psi^{LC}_{qqq/N}(x_i, k_{\perp i},
\lambda_i)$ satisfying $SU(6)$ spin-flavor symmetry which can account for
the ``static" properties of the baryons: their magnetic moments, axial
couplings $g_A$, and charged radii.
\cite{Brodsky:1994fz,Schlumpf:1995ik} Such LC models satisfy the
rigorous constraint that the magnetic moment of a composite spin-half
state must approach its Dirac moment $\mu = e/2M$ in the pointlike limit
$R \to 0$ with $M$ fixed, where $R^2 = d F_1(q^2)/d q^2\vert_{q^2 \to
0}$.  In addition, the LC model predicts that the quark chirality measures
$\Delta q, \Delta \Sigma,$ and $ g_A$ vanish in the same pointlike
limit.  For the physical proton, their values are approximately $0.75$ of
the nonrelativistic values.  Physically, this reduction occurs because
the quark chirality, (which can be identified with quark helicity $\vec
S_q \cdot \hat p$ in the massless limit) fluctuates strongly as the bound
state becomes pointlike.  Thus one cannot identify the chirality measures
which appear in the Bjorken and Ellis-Jaffe-Gourdin sum rules in a
relativistic theory with the spin projection of the equal-time
wavefunction in the hadron rest-frame.  One can also construct exact models
based on the perturbative structure of the QED calculation of the
anomalous moment \cite{BD} using Pauli-Villars spectra.\cite{BHMS}
As discussed by Bo-Qiang Ma in these
proceedings, the chirality sum rules are effectively measures of the
light-cone spin projections, not the usual equal-time spin.\cite{Ma:1999wk}

\section{Intrinsic versus Extrinsic Sea}

The deep inelastic scattering data show that the
nonperturbative structure of nucleons is more complex than a simple three
quark bound state.  For example, if the sea quarks were generated solely
by perturbative QCD evolution via gluon splitting, the anti-quark
distributions would be approximately isospin symmetric.  However, the
$\bar u(x)$ and
$\bar d(x)$ antiquark distributions of the proton at $Q^2
\sim 10$ GeV$^2$ are found to be quite different in
shape  \cite{{Nasalski:1994bh}}
and thus must reflect dynamics intrinsic to the proton's structure.
Evidence for a difference between the $\bar s(x)$ and $s(x)$ distributions
has also been claimed.\cite{Barone:1999yv} There have also been surprises
associated with the chirality distributions
$\Delta q = q_{\uparrow/\uparrow} - q_{\downarrow/\uparrow}$ of the valence
quarks which show that a simple valence quark
approximation to nucleon spin structure functions is far from the actual
dynamical situation.\cite{Karliner:1999fn}

It is helpful to categorize the parton distributions as
``intrinsic"---pertain\-ing to
the long-time scale composition of the target hadron, and
``extrinsic"---reflecting
the short-time substructure of the individual quarks and gluons themselves.
Gluons carry a
significant fraction of the proton's spin as well as its momentum.  Since
gluon exchange between valence quarks contributes to the
$p-\Delta$ mass splitting, it follows that the gluon distributions
cannot be solely accounted for by gluon bremsstrahlung from
individual quarks, the
process responsible for DGLAP evolutions of the structure functions.
Similarly,  in the case of heavy quarks, $s\bar s$,
$c \bar c$, $b \bar b$, the diagrams in which the sea quarks are
multi-connected to
the valence quarks are intrinsic to the proton structure itself.\cite{IC}

The higher Fock state of the proton $\ket{u u d s \bar s}$ should
resemble a $\ket{ K \Lambda}$ intermediate state, since this minimizes its
invariant mass $\M$.  In such a state, the
strange quark has a higher mean momentum fraction $x$ than the $\bar
s$.\cite{Warr,Signal,BMa} Similarly, the helicity intrinsic strange
quark in this configuration will be anti-aligned with the helicity of the
nucleon.\cite{Warr,BMa} This $Q \leftrightarrow \bar Q$ asymmetry
is a striking feature of the intrinsic heavy-quark sea.

In a recent paper, Merino, Rathsman, and I have shown that the asymmetry in the
fractional energy of charm versus anticharm jets produced in high energy
diffractive photoproduction is sensitive to the interference of the Odderon
$(C = -)$ and Pomeron
$(C = +)$ exchange amplitudes in QCD.  We can predict the dynamical shape
of the asymmetry in a simple model and have estimated its magnitude to be
of the order 15\% using an Odderon coupling to the proton which
saturates constraints from proton-proton vs.~proton-antiproton elastic
scattering.  Measurements of this asymmetry
at HERA could provide the first evidence for the presence of Odderon
exchange in the high energy limit of strong interactions.

The main features of the heavy sea quark-pair contributions of the Fock
state expansion of light hadrons can be derived from perturbative QCD,
since $\M^2_n$ grows with
$m^2_Q$.  One identifies two contributions to the heavy quark sea, the
``extrinsic'' contributions which correspond to ordinary gluon splitting, and
the ``intrinsic" sea which is multi-connected via gluons to the valence quarks.
The intrinsic sea is thus sensitive to the hadronic bound state
structure.\cite{IC} The maximal contribution of the
intrinsic heavy quark occurs at $x_Q \simeq {m_{\perp Q}/ \sum_i m_\perp}$
where $m_\perp = \sqrt{m^2+k^2_\perp}$;
\ie\ at large $x_Q$, since this minimizes the invariant mass $\M^2_n$.
The
measurements of the charm structure function by the EMC experiment are
consistent with intrinsic charm at large $x$ in the nucleon with a
probability of order $0.6 \pm 0.3 \% $.\cite{Harris:1996jx}
Similarly, one can distinguish intrinsic
gluons which are associated with multi-quark interactions and extrinsic gluon
contributions associated with quark substructure.\cite{BS} One can also
use this
framework to isolate the physics of the anomaly contribution to the Ellis-Jaffe
sum rule.\cite{Bass:1998rn} Thus neither gluons nor sea
quarks are solely generated by DGLAP evolution, and one cannot define a
resolution scale $Q_0$ where the sea or gluon degrees of freedom can be
neglected.

Light-cone wavefunctions are the natural quantities to encode hadron
properties and to bridge the gap between empirical constraints and
theoretical predictions for the bound state solutions.  We can thus
envision a program to construct the $\Psi^P_n(x_i, k_{\perp i},
\lambda_i)$ using not only data, but theoretical constraints such as

(1) Since the state is far off shell at large invariant mass $\M$,
one can derive rigorous limits on the
$x \to 1$, high $k_\perp$, and high
$\M^2_n$ behavior of the wavefunctions in the perturbative
domain.\cite{LB,Hoyer:1990pa}

(2) Ladder relations connecting state of different particle number
follow from the QCD equation of motion and lead to Regge behavior of the
quark and gluon distributions at $x \to 0$.  QED provides a constraint at
$N_C \to 0.$ \cite{ABD}

(3) One can obtain guides to the exact behavior of LC wavefunctions in
QCD from analytic or DLCQ solutions to toy models such as ``reduced"
$QCD(1+1).$ \cite{AD}

(4) QCD sum rules, lattice gauge theory moments, and QCD inspired models
such as the bag model, chiral theories, provide important constraints.

(5) Since the LC formalism is valid at all scales, one can utilize
empirical constraints such as the measurements of magnetic
moments, axial couplings, form factors, and distribution amplitudes.

(6) In the nonrelativistic limit, the light-cone and many-body
Schr\"odinger theory formalisms must match.

\section{ The Light-Cone Factorization Scheme}

Factorization theorems for hard exclusive, semi-exclusive, and
diffractive processes allow a rigorous separation of soft
non-perturbative dynamics of the bound state hadrons from the hard
dynamics of a perturbatively-calculable quark-gluon scattering
amplitude.

Roughly, the direct proofs of factorization in the light-cone scheme
proceed as follows: \cite{LB} In hard inclusive reactions all intermediate
states are divided according to $\M^2_n < \Lambda^2 $ and $\M^2_n >
\Lambda^2 $ domains.  The lower region is associated with the quark and
gluon distributions defined from the absolute squares of the LC
wavefunctions in the light cone factorization scheme.  In the high
invariant mass regime, intrinsic transverse momenta can be ignored, so
that the structure of the process at leading power has the form of hard
scattering on collinear quark and gluon constituents, as in the parton
model.  The attachment of gluons from the LC wavefunction to a propagator
in the hard subprocess is power-law suppressed in LC gauge, so that the
minimal
$2 \to 2$ quark-gluon subprocesses dominate.  The
higher order loop corrections lead to the DGLAP evolution equations, as well as
the higher order in $\alpha_s$ corrections to the hard amplitude.

It is important to note that the
effective starting point for the PQCD evolution of the structure
functions cannot be taken as a constant
$Q^2_0$ since as
$x \to 1$ the invariant mass $\M_n$ exceeds the resolution scale
$\Lambda$.  Thus in effect, evolution
is quenched at $ x \to 1$.\cite{LB,BrodskyLepage,Dmuller}

One of the most interesting aspects of deep inelastic lepton-proton
scattering is the contribution to the $g_1^p$ spin-dependent structure
function from photon-gluon fusion subprocesses $\gamma^*(q)
g(p)\rightarrow q\bar q$.\cite{Carlitz:1988ab,Bass:1998rn} Naively, one
would expect zero contributions from light mass $q\bar q$ pairs to the
first moment $\int^1_0 dx\, g_1^p(x,Q^2)$ since the $q$ and $\bar q$ have
opposite helicities.  In fact, this is not the case if the quark mass
$m_q$ is small compared to a scale set by the spacelike gluon virtuality
$p^2$.  This is the origin of the so-called anomalous correction
$-3\frac{\alpha_s}{2\pi}\, \Delta g$ to the Ellis-Jaffe sum rule for
isospin zero targets assuming three light flavors.  Here $\Delta g$ is
the helicity carried by gluons in the hadron target, $\Delta g(Q) =
\int^1_0 dx [g_\uparrow (x,Q) - g_\downarrow(x,Q)]$, at the
factorization scale $Q$.
If the sea quark mass is heavy compared to the gluon
virtuality $4m^2_q \gg P^2 = - p^2$, the photon-gluon fusion
contribution to $\int_0^1 dx\, g_1(x,Q^2)$ vanishes to leading order in
$\alpha_s(Q^2)$.  This result follows from a general theorem based on
the Drell-Hearn-Gerasimov sum rule which states that the
integral
\begin{equation}
\int ^\infty_{\nu_\pi} \frac{d\nu}{\nu}\
\sigma_{\gamma a\rightarrow bc}(\nu) = 0(\alpha^3) \ ;
\label{eq:2}
\end{equation}
\ie, vanishes at order $\alpha^2$ for any $2\rightarrow 2$ Standard
Model process.\cite{Alt72,Brod95} In the present case the gluon (for
$p^2 = 0$) takes the role of the target $a$.  For large $Q^2$, the DHG
integral evolves to the first moment of the helicity-dependent structure
function $g_1(x,Q^2)$ for any photon virtuality.  Thus the fusion
$\gamma^* g\rightarrow q\bar q$ contribution to $\int_0^1 dx\,
g_1(x,Q^2)$ vanishes for small gluon virtuality $P^2 \ll 4m^2_q$, $P^2
\ll Q^2$.  This virtuality can be interpreted directly in the
light-cone factorization scheme.  If the off-shellness of the state
is larger than the quark pair mass, one obtains the usual anomaly
contribution.\cite{Carlitz:1988ab,Bass:1998rn} The specific
contribution of a given sea quark pair $q\bar q$ thus depends not only on
$Q^2$, but more critically on the ratio of scales $p^2/4m^2_q$.  The
spectrum $N(p^2)$ of gluon virtuality in the target nucleon in turn
depends in detail on the physics of the nucleon light-cone wavefunction.
Bass, Schmidt and I\cite{Bass:1998rn} have discussed specific forms which
allow one to estimate the effect of extrinsic and intrinsic $s$ and $c$
quarks on the anomaly.  The application of the DHG theorem to
photoabsorption is more general than leading twist.\cite{bassbs} The
fusion contribution to the DHG moment vanishes even if $Q^2 < 4m^2_q$,
as long as the gluon virtuality can be neglected.  The result also holds
for the weak as well as electromagnetic current probes.\cite{Brod95,rizzo}

In exclusive amplitudes, the LC wavefunctions are the interpolating
functions between the quark and gluon states and the hadronic states.
In an
exclusive amplitude involving a hard scale $Q^2$, the intermediate states
can again be divided in invariant mass domains.  The
high invariant mass contributions to the amplitude has the structure of a
hard scattering process
$T_H$ in which the hadrons are replaced by their respective (collinear)
quarks and gluons.  In light-cone gauge only the minimal Fock states
contribute to the leading power-law fall-off of the exclusive amplitude.
The wavefunctions in the lower invariant mass domain can be integrated up
to the invariant mass cutoff $\Lambda$ and replaced by the gauge
invariant distribution amplitudes, $\phi_H(x_i,\Lambda)$.  Final-state
and initial-state corrections from gluon attachments to lines
connected to the color-singlet distribution amplitudes cancel at
leading twist.  Thus the key non-perturbative input for exclusive
processes is the gauge and frame independent hadron distribution
amplitude  \cite{LB} defined as the integral of the valence (lowest
particle number) Fock wavefunction;
\eg\ for the pion
\begin{equation}
\phi_\pi (x_i,\Lambda) \equiv \int d^2k_\perp\, \psi^{(\Lambda)}_{q\bar
q/\pi} (x_i, \vec k_{\perp i},\lambda)
\label{eq:f1}
\end{equation}
where the global cutoff $\Lambda$ is identified with the resolution $Q$.
The distribution amplitude controls leading-twist exclusive amplitudes
at high momentum transfer, and it can be related to the gauge-invariant
Bethe-Salpeter wavefunction at equal light-cone time.  The
logarithmic evolution of hadron distribution amplitudes
$\phi_H (x_i,Q)$ can be derived from the perturbatively-computable tail
of the valence light-cone wavefunction in the high transverse momentum
regime.\cite{LB}

The features of exclusive processes to leading power in the transferred
momenta are well known:

(1) The leading power fall-off is given by dimensional counting rules for
the hard-scattering amplitude: $T_H \sim 1/Q^{n-1}$, where $n$ is the total
number
of fields
(quarks, leptons, or gauge fields) participating in the hard
scattering.\cite{BF,Matveev:1973ra} Thus the reaction is dominated by
subprocesses
and Fock states involving the minimum number of interacting fields.  The
hadronic
amplitude follows this fall-off modulo logarithmic corrections from the
running of
the QCD coupling, and the evolution of the hadron distribution amplitudes.
In some
cases, such as large angle $p p \to p p $ scattering, pinch contributions from
multiple hard-scattering processes must also be
included.\cite{Landshoff:1974ew}
The general success of dimensional counting rules implies that the
effective coupling
$\alpha_V(Q^*)$ controlling the gluon exchange propagators in
$T_H$ are frozen in the infrared, \ie, have an infrared fixed point, since the
effective momentum transfers $Q^*$ exchanged by the gluons are often a
small fraction
of the overall momentum transfer.\cite{Brodsky:1998dh}
The pinch contributions
are then suppressed by a factor decreasing faster
than a fixed power.\cite{BF}

(2) The leading power dependence is given by hard-scattering amplitudes $T_H$
which conserve quark helicity.\cite{Brodsky:1981kj,Chernyak:1999cj} Since the
convolution of $T_H$ with the light-cone wavefunctions projects out states with
$L_z=0$, the leading hadron amplitudes conserve hadron helicity; \ie, the
sum of
initial and final hadron helicities are conserved.
Hadron helicity conservation
thus follows from the underlying chiral structure of QCD.  For example,
hadron helicity conservation predicts the suppression of vector meson
states produced with $J_z =\pm 1$ in $e^+ e^=$ annihilation to
vector-pseudoscalar final states.  However, $J/\psi \to \rho \pi$ appears
to occur copiously whereas
$\psi^\prime \to \rho \pi$ has never been conserved.  The PQCD analysis
assumes that
a heavy quarkonium state such as the
$J/\psi$ always decays to light hadrons via the annihilation of its heavy quark
constituents to gluons.  However, as Karliner and I \cite{Brodsky:1997fj}
have shown, the transition $J/\psi \to \rho
\pi$ can also occur by the rearrangement of the $c \bar c$ from the $J/\psi$
into the $\ket{ q \bar q c \bar c}$ intrinsic charm Fock state of the $\rho$ or
$\pi$.  On the other hand, the overlap rearrangement integral in the
decay $\psi^\prime \to \rho \pi$ will be suppressed since the intrinsic
charm Fock state radial wavefunction of the light hadrons will evidently
not have nodes in its radial wavefunction.  This observation can provide
a natural explanation of the long-standing puzzle why the $J/\psi$ decays
prominently to two-body pseudoscalar-vector final states, whereas the
$\psi^\prime$ does not.

I will mention here several other applications of the light-cone
formalism and factorization scheme:

{\it Diffractive vector meson photoproduction.} The
light-cone Fock wavefunction representation of hadronic amplitudes
allows a simple eikonal analysis of diffractive high energy processes, such as
$\gamma^*(Q^2) p \to \rho p$, in terms of the virtual photon and the vector
meson Fock state light-cone wavefunctions convoluted with the $g p \to g p$
near-forward matrix element.\cite{BGMFS} One can easily show that only small
transverse size $b_\perp \sim 1/Q$ of the vector meson distribution
amplitude is involved.  The hadronic interactions are minimal, and thus the
$\gamma^*(Q^2) N \to
\rho N$ reaction can occur coherently throughout a nuclear target in reactions
without absorption or shadowing.  The $\gamma^* A \to V A$ process
is thus a laboratory for testing QCD color transparency.\cite{BM}

{\it Regge behavior of structure functions.} The light-cone wavefunctions
$\psi_{n/H}$ of a hadron are not independent of each other, but rather are
coupled via the equations of motion.  The constraint of finite
``mechanical'' kinetic energy allows one to derive ``ladder relations"
which interrelate the light-cone wavefunctions of states differing by one
or two gluons.\cite{ABD} We can then use these relations to derive the
Regge behavior of both the polarized and unpolarized structure functions at $x
\rightarrow 0$, extending Mueller's derivation of the BFKL hard
QCD pomeron using the properties of heavy quarkonium light-cone
wavefunctions at large $N_C$ QCD.\cite{Mueller}

{\it Structure functions at large $x_{bj}$.} The behavior of structure functions
where one quark has the entire momentum requires the knowledge of LC
wavefunctions
with $x \rightarrow 1$ for the struck quark and $x \rightarrow 0$ for the
spectators.  This is a highly off-shell configuration, and thus one can
rigorously derive quark-counting and helicity-retention rules for the
power-law behavior of
the polarized and unpolarized quark and gluon distributions in the $x
\rightarrow 1$ endpoint domain.
Modulo DGLAP evolution, the counting rule for finding parton $a$ in hadron
$a$ at
large $x \sim 1$
$G_{a/A}(x,Q)
\propto (1-x)^{2 n_{\rm spect} - 1 + 2\vert
\Delta S_z\vert}$
where $n_{\rm spect}$ is the minimum number of partons left behind when parton
$a$ is
removed from $A$, and $\Delta S_z$ is the difference of the $a$ and
$A$ helicities.  This predicts $(1-x)^3$ behavior for valence quarks aligned in
helicity with the proton helicity, and $(1-x)^3$ behavior for anti-aligned
quarks.
As noted above, DGLAP evolution is quenched in the large $x$ limit in the fixed
$W^2$ domain.  Burkardt, Schmidt, and I have discussed the phenomenological
implications of this rule for gluon and sea distributions.

{\it Materialization of far-off-shell configurations.}
In a high energy hadronic collisions, the highly-virtual states of a hadron
can be
materialized into physical hadrons simply by the soft interaction of any of the
constituents.\cite{BHMT} Thus a proton state with intrinsic charm $\ket{ u
u d \bar c c}$ can be materialized by the
interaction of a light-quark in the target,
producing a $J/\psi$ at large $x_F$.  The production
occurs on the front-surface of a target nucleus, implying an $A^{2/3}$
$J/\psi$ production cross section at large $x_F,$ which is consistent
with experiment, such as Fermilab experiments E772 and E866.

{\it Comover phenomena.}
Light-cone wavefunctions describe not only the partons that interact in a hard
subprocess but also the associated partons freed from the projectile.  The
projectile partons which are comoving (\ie, which have similar rapidity) with
the final state quarks and gluons can interact strongly producing (a)
leading particle effects, such as those seen in open charm
hadroproduction; (b) suppression of quarkonium \cite{BrodskyMueller} in
favor of open heavy hadron production, as seen in the E772 experiment;
(c) changes in color configurations and selection rules in quarkonium
hadroproduction, as has been emphasized by Hoyer and
Peigne.\cite{Hoyer:1998ha} Further, more than one parton from the
projectile can enter the hard subprocess, producing dynamical higher twist
contributions, as seen for example in
Drell-Yan experiments.\cite{BrodskyBerger,Brandenburg}

{\it Jet hadronization in light-cone QCD.}
One of the goals of nonperturbative analysis in QCD is to compute jet
hadronization from first principles.  The DLCQ solutions provide a possible
method to accomplish this.  By inverting the DLCQ solutions, we can write the
``bare'' quark state of the free theory as
$\ket{q_0} = \sum \ket n \VEV{n\,|\,q_0}$
 where now $\{\ket n\}$ are the exact DLCQ eigen states of
$H_{LC}$, and
$\VEV{n\,|\,q_0}$ are the DLCQ projections of the eigen-solutions.  The
expansion
in automatically infrared and ultraviolet regulated if we impose global cutoffs
on the DLCQ basis:
$\lambda^2 < \Delta\M^2_n < \Lambda^2
$
where $\Delta\M^2_n = \M^2_n-(\Sigma \M_i)^2$.  It would be
interesting to study jet hadronization at the amplitude level for
the existing DLCQ solutions to QCD (1+1) and collinear QCD.

{\it Hidden Color.}
The deuteron form factor at high $Q^2$ is sensitive to wavefunction
configurations where all six quarks overlap within an impact
separation $b_{\perp i} < {\cal O} (1/Q);$ the leading power-law
fall off predicted by QCD is $F_d(Q^2) = f(\alpha_s(Q^2))/(Q^2)^5$,
where, asymptotically, $f(\alpha_s(Q^2)) \propto
\alpha_s(Q^2)^{5+2\gamma}$.\cite{Brodsky:1976rz} The derivation of the
evolution equation for the deuteron distribution amplitude and its
leading anomalous dimension $\gamma$ is given by
Ji, Lepage, and myself.~\cite{bjl83}
In general, the six-quark wavefunction of a deuteron
is a mixture of five different color-singlet states.  The dominant
color configuration at large distances corresponds to the usual
proton-neutron bound state.  However at small impact space
separation, all five Fock color-singlet components eventually
acquire equal weight, \ie, the deuteron wavefunction evolves to
80\%\ ``hidden color.'' The relatively large normalization of the
deuteron form factor observed at large $Q^2$ points to sizable
hidden color contributions.\cite{Farrar:1991qi} Hidden color components
can play a predominant role in the reaction $\gamma d \to J/\psi p n$ at
threshold if it is dominated by the multi-fusion process $\gamma g g \to
J/\psi$.

{\it Spin-Spin Correlations and the Charm Threshold.}
One of the most striking anomalies in elastic proton-proton
scattering is the large spin correlation $A_{NN}$ observed at large
angles.\cite{krisch92} At $\sqrt s \simeq 5 $ GeV, the rate for
scattering with incident proton spins parallel and normal to the
scattering plane is four times larger than that for scattering with
anti-parallel polarization.  This strong polarization correlation can
be attributed to the onset of charm production in the intermediate
state at this energy.\cite{Brodsky:1988xw,deTeramond:1998ny}
A resonant intermediate state $\vert u
u d u u d c \bar c \rangle$ has odd intrinsic parity and can thus couple
to the $J=L=S=1$ initial state, thus strongly enhancing scattering when
the incident projectile and target protons have their spins parallel
and normal to the scattering plane.  The charm threshold can also
explain the anomalous change in color transparency observed at the
same energy in quasi-elastic $ p p$ scattering.  A crucial test is
the observation of open charm production near threshold with a
cross
section of order of $1 \mu$b.  Analogous strong spin effects should also
appear at the strangeness threshold and in exclusive photon-proton
reactions such as large angle Compton scattering and pion photoproduction
near the strangeness and charm thresholds.
\section{Self-Resolved Diffractive Reactions and Light Cone Wavefunctions}

Diffractive multi-jet production in heavy
nuclei provides a novel way to measure the shape of the LC Fock
state wavefunctions and test color transparency.  For example, consider the
reaction  \cite{Bertsch,MillerFrankfurtStrikman,Frankfurt:1999tq}
$\pi A \rightarrow {\rm Jet}_1 + {\rm Jet}_2 + A^\prime$
at high energy where the nucleus $A^\prime$ is left intact in its ground
state.  The transverse momenta of the jets have to balance so that
$
\vec k_{\perp i} + \vec k_{\perp 2} = \vec q_\perp < {R^{-1}}_A \ ,
$
and the light-cone longitudinal momentum fractions have to add to
$x_1+x_2 \sim 1$ so that $\Delta p_L < R^{-1}_A$.  The process can
then occur coherently in the nucleus.  Because of color transparency,  \ie,
the cancelation of color interactions in a small-size color-singlet
hadron,  the valence wavefunction of the pion with small impact
separation, will penetrate the nucleus with minimal interactions,
diffracting into jet pairs.\cite{Bertsch}
The $x_1=x$, $x_2=1-x$ dependence of
the di-jet distributions will thus reflect the shape of the pion distribution
amplitude; the $\vec k_{\perp 1}- \vec k_{\perp 2}$
relative transverse momenta of the jets also gives key information on
 the underlying shape of the valence pion
wavefunction.\cite{MillerFrankfurtStrikman,Frankfurt:1999tq} The QCD
analysis can be
confirmed by the observation that the diffractive nuclear amplitude
extrapolated to
$t = 0$ is linear in nuclear number $A$, as predicted by QCD color
transparency.  The integrated diffractive rate should scale as $A^2/R^2_A \sim
A^{4/3}$.  A diffractive dissociation experiment of this type, E791,  is now in
progress at Fermilab using 500 GeV incident pions on nuclear
targets.\cite{E791} The preliminary results from E791 appear to be consistent
with color transparency.  The momentum fraction distribution of the jets is
consistent with a valence light-cone wavefunction of the pion consistent with
the shape of the asymptotic distribution amplitude, $\phi^{\rm asympt}_\pi (x) =
\sqrt 3 f_\pi x(1-x)$.  Data from
CLEO \cite{Gronberg:1998fj} for the
$\gamma
\gamma^* \rightarrow \pi^0$ transition form factor also favor a form for
the pion distribution amplitude close to the asymptotic solution \cite{LB}
to the perturbative QCD evolution
equation.\cite{Kroll,Rad,Brodsky:1998dh,Feldmann:1999wr,Schmedding:1999ap}
It will also be interesting to study diffractive tri-jet production using proton
beams
$ p A \rightarrow {\rm Jet}_1 + {\rm Jet}_2 + {\rm Jet}_3 + A^\prime $ to
determine the fundamental shape of the 3-quark structure of the valence
light-cone wavefunction of the nucleon at small transverse
separation.\cite{MillerFrankfurtStrikman} One interesting possibility is
that the distribution amplitude of the
$\Delta(1232)$ for $J_z = 1/2, 3/2$ is close to the asymptotic form $x_1
x_2 x_3$,  but that the proton distribution amplitude is more complex.
This would explain why the $p \to\Delta$ transition form factor appears to
fall faster at large $Q^2$ than the elastic $p \to p$ and the other $p \to
N^*$ transition form factors.\cite{Stoler:1999nj}
Conversely, one can use incident real and virtual photons:
$ \gamma^* A \rightarrow {\rm Jet}_1 + {\rm Jet}_2 + A^\prime $ to
confirm the shape of the calculable light-cone wavefunction for
transversely-polarized and longitudinally-polarized virtual photons.  Such
experiments will open up a direct window on the amplitude
structure of hadrons at short distances.

The diffractive dissociation of a hadron or nucleus can also occur via
the Coulomb dissociation of a beam particle on an electron beam (\eg\ at
HERA or eRHIC) or on the strong Coulomb field of a heavy nucleus (\eg\
at RHIC or nuclear collisions at the LHC).\cite{BHDP} The amplitude for
Coulomb exchange at small momentum transfer is proportional to the first
derivative $\sum_i e_i {\partial \over \vec k_{T i}} \psi$ of the
light-cone wavefunction, summed over the charged constituents.  The Coulomb
exchange reactions fall off less fast at high transverse momentum compared
to pomeron exchange reactions since the light-cone wavefunction is
effective differentiated twice in two-gluon exchange reactions.

For example, consider the Coulomb dissociation of a high energy proton at
HERA.  The proton can dissociate into three jets corresponding to the
three-quark structure of the valence light-cone wavefunction.  We can
demand that the produced hadrons all fall outside of an ``exclusion cone"
of opening angle $\theta$ in the proton's fragmentation region.
Effectively all of the light-cone momentum
$\sum_j x_j \simeq 1$ of the proton's fragments will thus be
produced outside the exclusion cone.  This requirement
then limits the invariant mass of the Fock state ${\cal M}^2_n >
\Lambda^2 = P^{+2} \sin^2\theta/4$ from below, so that perturbative QCD
counting rules can predict the fall-off in the jet system invariant mass
$\cal M$.  At large invariant mass one expects the three-quark valence
Fock state of the proton to dominate.  The segmentation of the forward
detector in azimuthal angle $\phi$ can be used to identify structure and
correlations associated with the three-quark light-cone wavefunction.
A further discussion is in progress.~\cite{BHDP}

The light-cone formalism is also applicable to the
description of nuclei in terms of their nucleonic and mesonic
degrees of freedom.\cite{Miller:1999mi}
Self-resolving diffractive jet reactions
in high energy electron-nucleus collisions and hadron-nucleus collisions
at moderate momentum transfers can thus be used to resolve the light-cone
wavefunctions of nuclei.

\section{Semi-Exclusive Processes:  New Probes of Hadron Structure}

A new class of hard ``semi-exclusive''
processes of the form $A+B \to C + Y$, have been proposed as new probes of
QCD.\cite{BB,acw,Brodsky:1998sr} These processes are characterized
by a large momentum transfer $t= (p_A-p_C)^2$ and a large rapidity gap between
the final state particle $C$ and the inclusive system $Y$.
Here $A, B$
and $C$ can be hadrons or (real or virtual) photons.  The cross
sections for such processes factorize in terms of the distribution
amplitudes of $A$ and $C$ and the parton distributions in the target
$B$.  Because of this factorization, semi-exclusive reactions provide a
novel array of generalized currents, \cite{Brodsky:1998sr} which not only
give insight into the dynamics of hard scattering QCD processes, but also
allow experimental access to new combinations of the universal quark and
gluon distributions.

\section{Summary}

In this talk I have
discussed how universal, process-independent and
frame-independent light-cone Fock-state wavefunctions can be used to
encode the properties of a hadron in terms of its fundamental quark and
gluon degrees of freedom.  Given the proton's light-cone wavefunctions,
one can compute not only the moments of the quark and gluon distributions
measured in deep inelastic lepton-proton scattering, but also the
multi-parton correlations which control the distribution of particles in
the proton fragmentation region and dynamical higher twist effects.
Light-cone wavefunctions also provide a systematic framework for
evaluating exclusive hadronic matrix elements, including time-like heavy
hadron decay amplitudes and form factors.  The formalism also provides a
physical factorization scheme for separating hard and soft contributions
in both exclusive and inclusive hard processes.  A new type of jet
production reaction, ``self-resolving diffractive interactions" can
provide direct information on the light-cone wavefunctions of hadrons in
terms of their QCD degrees of freedom, as well as the composition of
nuclei in terms of their nucleon and mesonic degrees of freedom.
Progress in QCD is driven by experiment, and we are fortunate that there
are new experimental facilities such as Jefferson laboratory, new
studies of exclusive processes $e^+ e^-$ and $\gamma \gamma$ processes at
the high
luminosity
$B$ factories, as well as the new accelerators and colliders now being
planned to
further advance the study of QCD phenomena.

\section*{Acknowledgments}
Work supported by the Department of Energy
under contract number DE-AC03-76SF00515.
I wish to thank Shunzo Kumano and Akihisa Kohama
for their kind hospitality at this symposium and the RIKEN
Institute.  I also
thank Paul Hoyer, Markus Diehl, Stephane Peigne, Bo-Qiang Ma, Ivan
Schmidt, and Dae Sung Hwang for helpful conversations.


\begin{thebibliography}{99}



\bibitem{Coward:1968au}
D.~H.~Coward {\it et al.},
Phys.\ Rev.\ Lett.\  {\bf 20}, 292 (1968).


\bibitem{Dirac:1949cp}
P.~A.~Dirac,
Rev.\ Mod.\ Phys.\  {\bf 21}, 392 (1949).

\bibitem{PinskyPauli}
For a review and further references see S. J.~Brodsky, H.~Pauli and S.
S.~Pinsky, {\em Phys. Rept.}  {\bf 301} (1998) 299, hep-ph/9705477.



\bibitem{Srivastava:1999gi}
P. P.~Srivastava and S. J.~Brodsky, hep-ph/9906423.

\bibitem{LB}
G. P. Lepage and S. J. Brodsky, {\em Phys. Rev.} {\bf D22} (1980) 2157;
{\em Phys. Lett.} {\bf B87}, 359 (1979); {\em Phys. Rev. Lett.} {\bf 43}
(1979) 545, 1625(E).


\bibitem{BM}
S. J. Brodsky and A. H. Mueller, {\em Phys. Lett.} {\bf 206B} (1988) 685;
  L. Frankfurt and M. Strikman, {\em Phys. Rept.} {\bf  160} (1988), {235};
  P. Jain, B. Pire and J. P. Ralston, {\em Phys. Rept.} {\bf 271} (1996) {67}.

\bibitem{BD}
S. J. Brodsky and S. D. Drell, {\em Phys. Rev.} {\bf D22} (1980) 2236.

\bibitem{Brodsky:1998hn}
S. J.~Brodsky and D. S.~Hwang,
{\em Nucl.\ Phys.}\ {\bf B543} (1999) 239,
hep-ph/9806358.

\bibitem{CRY}
S. J. Chang, R. G. Root and T. M. Yan, {\em Phys. Rev.} {\bf D7} (1973) 1133.

\bibitem{BUR}
M. Burkardt, {\em Nucl. Phys.} {\bf A504} (1989) 762;
{\em Nucl. Phys.} {\bf B373} (1992) 613; {\em Phys. Rev.} {\bf D52} (1995) 3841.

\bibitem{Choi:1998nf}
H.~Choi and C.~Ji,
{\em Phys.\ Rev.}\ {\bf D58} (1998) 071901, hep-ph/9805438.

\bibitem{Pauli:1985ps}
H.~C.~Pauli and S.~J.~Brodsky,
Phys.\ Rev.\  {\bf D32}, 2001 (1985).



\bibitem{Kleb}
S. Dalley, and I. R. Klebanov, {\em Phys. Rev.} {\bf D47} (1993) 2517.

\bibitem{AD}
F. Antonuccio and S. Dalley, {\em Phys. Lett.} {\bf B348} (1995) 55;
{\em Phys. Lett.} {\bf B376}, 154 (1996); {\em Nucl. Phys.} {\bf B461}
(1996) 275.

\bibitem{Brodsky:1998hs}
S. J.~Brodsky, J. R.~Hiller and G.~McCartor,
{\em Phys.\ Rev.}\ {\bf D58}, 025005 (1998)
hep-th/9802120.


\bibitem{Brodsky:1999xj}
S.~J.~Brodsky, J.~R.~Hiller and G.~McCartor,
{\em Phys.\ Rev.}\  {\bf D60}, 054506 (1999)
[hep-ph/9903388].

\bibitem{Dalley:1999ii}
S.~Dalley and B.~van de Sande,
hep-lat/9911035.



\bibitem{Lunin:1999ib}
O.~Lunin and S.~Pinsky,
hep-th/9910222.



\bibitem{Haney:1999tk}
P.~Haney, J.~R.~Hiller, O.~Lunin, S.~Pinsky and U.~Trittmann,
hep-th/9911243.

\bibitem{Antonuccio:1999ia}
F.~Antonuccio, I.~Filippov, P.~Haney, O.~Lunin, S.~Pinsky, U.~Trittmann and
J.~Hiller
                  [SDLCQ Collaboration],
hep-th/9910012.

\bibitem{Bassetto:1999tm}
A.~Bassetto, L.~Griguolo and F.~Vian,
hep-th/9911036.


\bibitem{Brodsky:1994fz}
S.~J.~Brodsky and F.~Schlumpf,
{\em Phys.\ Lett.}\  {\bf B329}, 111 (1994)
[hep-ph/9402214].

\bibitem{Schlumpf:1995ik}
F.~Schlumpf and S.~J.~Brodsky,
{\em Phys.\ Lett.}\  {\bf B360}, 1 (1995)
[hep-ph/9505276].


\bibitem{BHMS}
S.~J.~Brodsky, D.~S. Hwang, B.-M.~Hwang, and I.~A.~Schmidt, in
preparation.

\bibitem{Ma:1999wk}
B.~Ma,
hep-ph/0001094.


\bibitem{Nasalski:1994bh}
J. P.~Nasalski [New Muon Collaboration],
{\em Nucl.\ Phys.}\ {\bf A577} (1994) 325C.

\bibitem{Barone:1999yv}
V.~Barone, C.~Pascaud and F.~Zomer, hep-ph/9907512.

\bibitem{Karliner:1999fn}
M.~Karliner and H. J.~Lipkin, hep-ph/9906321.

\bibitem{IC}
S. J. Brodsky, P. Hoyer, C. Peterson, and N. Sakai, {\em Phys. Lett.} {\bf
93B} (1980)  451.

\bibitem{Warr}
M. Burkardt and  Brian Warr, {\em Phys. Rev.} {\bf D45} (1992) 958.

\bibitem{Signal}
A. I. Signal and A. W. Thomas, {\em Phys. Lett.} {\bf 191B} (1987) 205.


\bibitem{BMa}
S. J. Brodsky and B.-Q. Ma, {\em Phys. Lett.} {\bf B381}, 317 (1996),
hep-ph/9604393.

\bibitem{Harris:1996jx}
B. W.~Harris, J.~Smith and R.~Vogt,
{\em Nucl.\ Phys.}\ {\bf B461} (1996) 181,
hep-ph/9508403.

\bibitem{BS}
S. J. Brodsky and I. A. Schmidt, {\em Phys. Lett.} {\bf B234} (1990) 144.

\bibitem{Bass:1998rn}
S.D.~Bass, S.J.~Brodsky and I.~Schmidt,
{\em Phys.\ Rev.}\ {\bf D60} (1999) 034010,
hep-ph/9901244.


\bibitem{Hoyer:1990pa}
P.~Hoyer and S.~J.~Brodsky,
SLAC-PUB-5374
{\it Invited talk given at Topical Conf. on Particle Production
near Threshold, Nashville, IN, Sep 30 - Oct 3, 1990}.

\bibitem{ABD}
F. Antonuccio,  S. J. Brodsky, and S. Dalley,
{\em Phys. Lett.} {\bf B412} (1997) 104, hep-ph/9705413.

\bibitem{BrodskyLepage}
S. J. Brodsky and G. P. Lepage, in {\em Perturbative Quantum
Chromodynamics}, A. H. Mueller, Ed.  (World Scientific, 1989).

\bibitem{Dmuller}
D. Mueller, SLAC-PUB-6496, May 1994,   hep-ph/9406260.


\bibitem{Carlitz:1988ab}
R.~D.~Carlitz, J.~C.~Collins and A.~H.~Mueller,
{\em Phys.\ Lett.}\  {\bf B214}, 229 (1988).


\bibitem{Alt72}
G.~Altarelli, N.~Cabibbo and L.~Maiani,
{\em Phys.\ Lett.}\  {\bf B40}, 415 (1972).

\bibitem{Brod95}
S.~J.~Brodsky and I.~Schmidt,
{\em Phys.\ Lett.}\  {\bf B351}, 344 (1995)
[hep-ph/9502416].

\bibitem{bassbs}
S.~D.~Bass, S.~J.~Brodsky and I.~Schmidt,
{\em Phys.\ Rev.}\  {\bf D60}, 034010 (1999)
[hep-ph/9901244].

\bibitem{rizzo}
S.~J.~Brodsky, T.~G.~Rizzo and I.~Schmidt,
{\em Phys.\ Rev.}\  {\bf D52}, 4929 (1995)
[hep-ph/9505441].

\bibitem{BF}
S. J.~Brodsky and G. R.~Farrar,
{\em Phys. Rev.} {\bf D11} (1975) 1309.

\bibitem{Matveev:1973ra}
V. A.~Matveev, R. M.~Muradian and A. N.~Tavkhelidze,
{\em Nuovo Cim. Lett.} {\bf 7} (1973) 719.

\bibitem{Landshoff:1974ew}
P. V.~Landshoff, {\em Phys. Rev.} {\bf D10} (1974) 1024.

\bibitem{Brodsky:1998dh}
S. J.~Brodsky, C.~Ji, A.~Pang and D. G.~Robertson,
{\em Phys.\ Rev.}\ {\bf D57} (1998) 245, hep-ph/9705221.

\bibitem{Brodsky:1981kj}
S. J.~Brodsky and G. P.~Lepage,
{\em Phys. Rev.} {\bf D24} (1981) 2848.

\bibitem{Chernyak:1999cj}
V.~Chernyak, hep-ph/9906387.

\bibitem{Brodsky:1997fj}
S. J.~Brodsky and M.~Karliner, {\em Phys. Rev. Lett.} {\bf 78} (1997) 4682,
hep-ph/9704379.

\bibitem{BGMFS}
S. J. Brodsky, L. Frankfurt, J. F. Gunion, A. H.
Mueller, and M. Strikman,  {\em Phys. Rev.} {\bf D50} (1994) 3134,
hep-ph/9402283.



\bibitem{Mueller}
A.  H.  Mueller,  {\em Phys. Lett.}  {\bf B308} (1993) 355.


\bibitem{BHMT}
S. J. Brodsky, P. Hoyer, A. H. Mueller, W.-K. Tang, {\em Nucl. Phys.} {\bf
B369} (1992) 519.

\bibitem{BrodskyMueller}
See S. J. Brodsky \etal.~\cite{BM}
R. Vogt, S. J. Brodsky, and P. Hoyer, {\em  Nucl. Phys.} {\bf B360} (1991) 67;
{\em  Nucl. Phys.} {\bf B383} (1992) 643.

\bibitem{Hoyer:1998ha}
P.~Hoyer and S.~Peigne, {\em Phys.\ Rev.}\ {\bf D59} (1999) 034011,
hep-ph/9806424.

\bibitem{BrodskyBerger}
E. L. Berger and S. J. Brodsky, {\em  Phys. Rev. Lett.} {\bf 42} (1979)  940.

\bibitem{Brandenburg}
A. Brandenburg, S. J. Brodsky, V.V. Khoze, and D. Mueller, {\em Phys. Rev.
Lett.} {\bf 73} (1994) 939, hep-ph/9403361.

\bibitem{Brodsky:1976rz}
S. J.~Brodsky and B. T.~Chertok, {\em Phys. Rev.} {\bf D14} (1976) 3003.

\bibitem{bjl83}
S. J. Brodsky, C.-R. Ji, and G. P. Lepage, {\em Phys. Rev. Lett.} {\bf
51} (1983) 83.

\bibitem{Farrar:1991qi}
G. R.~Farrar, K.~Huleihel and H.~Zhang,
{\em Phys. Rev. Lett.} {\bf 74} (1995) 650.

\bibitem{krisch92}
A. D. Krisch, {\em Nucl. Phys. B (Proc. Suppl.)} {\bf 25} (1992) 285.

\bibitem{Brodsky:1988xw}
S. J.~Brodsky and G. F.~de Teramond,
{\em Phys. Rev. Lett.} {\bf 60} (1988) 1924.

\bibitem{deTeramond:1998ny}
G.~F.~de Teramond, R.~Espinoza and M.~Ortega-Rodriguez,
{\em Phys.\ Rev.}\  {\bf D58}, 034012 (1998)
[hep-ph/9708202].

\bibitem{Bertsch}
G. Bertsch, S. J. Brodsky,
A. S. Goldhaber, and J. F. Gunion, {\em Phys. Rev. Lett.} {\bf 47} (1981) 297.

\bibitem{MillerFrankfurtStrikman}
L. Frankfurt, G. A. Miller, and M. Strikman, {\em Phys. Lett.}
{\bf B304} (1993) 1,  hep-ph/9305228.

\bibitem{Frankfurt:1999tq}
L.~Frankfurt, G. A.~Miller and M.~Strikman,
hep-ph/9907214.

\bibitem{E791}
D. F. Ashery \etal,  Fermilab E791 Collaboration, to be published.

\bibitem{Gronberg:1998fj}
J.~Gronberg {\em et al.} [CLEO Collaboration],
{\em Phys. Rev.} {\bf D57} (1998) 33,
hep-ex/9707031.

\bibitem{Kroll}
P. Kroll and M. Raulfs, {\em Phys. Lett.} {\bf B387} (1996) 848.

\bibitem {Rad}
I. V. Musatov and A. V. Radyushkin,  {\em Phys. Rev.} {\bf D56}
(1997) 2713.

\bibitem{Feldmann:1999wr}
T.~Feldmann, hep-ph/9907226.

\bibitem{Schmedding:1999ap}
A.~Schmedding and O.~Yakovlev, hep-ph/9905392.

\bibitem{Stoler:1999nj}
P.~Stoler,
{\em Few Body Syst.\ Suppl.}\ {\bf 11} (1999) 124.


\bibitem{BHDP}
S.~Brodsky,M.~Diehl,P.~Hoyer,S.~Peigne, in preparation.

\bibitem{Miller:1999mi}
G.~A.~Miller,
nucl-th/9910053.

\bibitem{BB}
  J. F. Gunion, S. J. Brodsky and R. Blankenbecler, {\em Phys. Rev.} {\bf
D6} (1972) {2652};
  R. Blankenbecler and S. J. Brodsky, {\em Phys. Rev.} {\bf D10} (1974) {2973}.

\bibitem{acw}
  C. E. Carlson and A. B. Wakely, {\em Phys. Rev.} {\bf D48} (1993) {2000};
  A. Afanasev, C. E. Carlson and C. Wahlquist, {\em Phys. Lett.} {\bf B398}
(1997) {393},
  hep-ph/9701215, and {\em Phys. Rev.} {\bf D58} (1998) {054007},
hep-ph/9706522.

\bibitem{Brodsky:1998sr}
S. J.~Brodsky, M.~Diehl, P.~Hoyer and S.~Peigne,
{\em Phys. Lett.} {\bf B449} (1999) 306, hep-ph/9812277.






\end{thebibliography}
\end{document}